\documentclass[fleqn,usenatbib]{mnras}

\usepackage[T1]{fontenc}

\DeclareRobustCommand{\VAN}[3]{#2}
\let\VANthebibliography\thebibliography
\def\thebibliography{\DeclareRobustCommand{\VAN}[3]{##3}\VANthebibliography}

\usepackage{graphicx}
\usepackage{amsmath}	
\usepackage{amssymb}	
\usepackage{bm}
\usepackage{hyperref}
\usepackage[normalem]{ulem}

\usepackage{newtxtext}
\usepackage[varvw]{newtxmath}

\newcommand{\diff}{\ensuremath{\mathrm{d}}}
\newcommand{\e}{\mathrm{e}}

\newcommand{\ac}{a_\mathrm{c}}
\newcommand{\dc}{\delta_\mathrm{c}}
\newcommand{\af}{a_\mathrm{f}}

\newcommand{\aeq}{a_\mathrm{eq}}

\newcommand{\keq}{k_\mathrm{eq}}
\newcommand{\Meq}{M_\mathrm{eq}}

\newcommand{\rhoM}{\rho_\mathrm{m}}
\newcommand{\rhoMo}{\rho_{\mathrm{m},0}}

\newcommand{\As}{A_\mathrm{s}}
\newcommand{\ns}{n_\mathrm{s}}

\def\chg#1{#1}

\title[Ultradense haloes accompany primordial black holes]{Ultradense dark matter haloes accompany primordial black holes}

\author[M. S. Delos \& J. Silk]{
M. Sten Delos,$^{1}$\thanks{E-mail: sten@mpa-garching.mpg.de}
Joseph Silk$^{2,3,4}$
\\
$^{1}$
Max Planck Institute for Astrophysics, Karl-Schwarzschild-Str. 1, 85748 Garching, Germany
\\
$^{2}$
Institut d'Astrophysique de Paris (UMR7095: CNRS \& UPMC- Sorbonne Universities), F-75014, Paris, France
\\
$^{3}$
Department of Physics and Astronomy, The Johns Hopkins University Homewood Campus, Baltimore, MD 21218, USA
\\
$^{4}$
BIPAC, Department of Physics, University of Oxford, Keble Road, Oxford OX1 3RH, UK
}

\date{Accepted XXX. Received YYY; in original form ZZZ}

\pubyear{2022}

\begin{document}

\label{firstpage}
\pagerange{\pageref{firstpage}--\pageref{lastpage}}
\maketitle

\begin{abstract}
Primordial black holes (PBHs) form from large-amplitude initial density fluctuations and may comprise some or all of the dark matter. If PBHs have a broadly extended mass spectrum, or in mixed PBH-particle dark matter scenarios, the extreme density fluctuations necessary to produce PBHs also lead to the formation of a much greater abundance of dark matter minihaloes that form during the radiation epoch with internal densities potentially of order $10^{12}$ M$_\odot$ pc$^{-3}$. We develop an analytic description of the formation of these ultradense haloes and use it to quantitatively compare PBH and halo distributions. PBHs that contribute only a per cent level fraction of the dark matter are accompanied by ultradense haloes that nevertheless comprise an order-unity fraction. \chg{These haloes would consist of either particle dark matter or much smaller PBHs.} This finding significantly alters the predictions of many PBH scenarios, enabling a variety of new observational tests.
\end{abstract}




\section{Introduction}

Primordial black holes (PBHs) are an attractive dark matter candidate \citep{2020ARNPS..70..355C,2021JPhG...48d3001G}. They form deep in the radiation-dominated epoch from large-amplitude initial density fluctuations, which occur on scales too small to influence precision cosmological observables like the large-scale galaxy distribution and temperature variations in the cosmic microwave background. In addition to the possibility that they comprise some or all of the dark matter, PBHs have been proposed to explain binary black hole coalescence detections by the LIGO/Virgo/KAGRA (LVK) collaboration \citep[e.g.][]{2016PhRvL.116t1301B,2016PhRvL.117f1101S} and to act as seeds for the supermassive black holes found at the centres of galaxies \citep[e.g.][]{2018MNRAS.478.3756C}.

An array of observational constraints significantly limit the possibility that PBHs in a narrow mass range comprise most of the dark matter \citep{2021RPPh...84k6902C}. However, extended PBH mass spectra remain viable, as do mixed scenarios with PBHs and particle dark matter. For example, while some properties of black hole systems seen in LVK observations \citep[e.g.][]{2020PhRvL.125j1102A} are difficult to realize with stellar black holes, they can be explained if PBHs around 10 to 100~M$_\odot$ contribute to the dark matter at the sub-per cent level \citep{2022PhRvD.105h3526F}. The same PBHs can also explain cored density profiles in dwarf galaxies \citep{2020MNRAS.492.5218B}. Similar mass fractions of much larger PBHs can seed supermassive black holes \citep[][]{2018MNRAS.478.3756C}, which are also challenging to explain through astrophysical formation channels \citep{2021NatRP...3..732V}. Extended PBH mass spectra that explain some or all of these phenomena can arise naturally from the timing of phase transitions in the early universe \citep{2018JCAP...08..041B,2021PDU....3100755C,2022arXiv220906196E}.

In mixed scenarios with PBHs and particle dark matter, and in scenarios where PBHs are all of the dark matter but have a sufficiently broad mass function, dark matter is present when PBHs are forming. Whereas PBHs form from extreme $\mathcal{O}(1)$ variations in the primordial density, weaker fluctuations that fail to form PBHs are far more common. These fluctuations are still sufficient to induce the formation of dark matter haloes long before they would be otherwise expected to form. Due to their early formation, these haloes would have extraordinarily high internal density. In particular, ultradense haloes can arise via the collapse of overdense regions deep in the radiation epoch.\footnote{Extremely compact haloes also grow \emph{around} PBHs \citep[e.g.][]{2009ApJ...707..979R,2019PhRvD.100h3528I} but represent a different phenomenon. The only link between the ultradense haloes that we discuss and PBHs is that they arise jointly from similar formation mechanisms.}

This \chg{article} develops an analytic description of ultradense halo formation and presents quantitative comparisons of PBH and ultradense halo populations. The idea that such haloes should vastly outnumber PBHs was considered previously by \citet{2009ApJ...707..979R} and \citet{2014PhRvD..90h3514K}, but the model of halo formation and structure employed therein disagrees with later numerical simulation results \citep{2017PhRvD..96l3519G,2018PhRvD..97d1303D,2018PhRvD..98f3527D}. These works also restricted their consideration to haloes forming during the last matter-dominated epoch; haloes that form during the radiation epoch can be far denser. \citet{1994PhRvD..50..769K}, \citet{2002JETP...94....1D}, and \citet{2010PhRvD..81j3529B,2013JCAP...11..059B} studied halo formation during the radiation epoch using models of spherical and ellipsoidal collapse. Motivated by a recent simulation result \citep{2019PhRvD.100j3010B}, we employ a related approach based on ellipsoidal collapse and the notion that halo formation proceeds once a collapsed region becomes locally matter dominated.

We show that haloes forming by a redshift of around $10^5$ can comprise an $\mathcal{O}(1)$ fraction of the dark matter even when PBHs arising from fluctuations at the same scale contribute only at the per cent level. Such early formation yields internal halo densities of order $10^{12}$~M$_\odot$\,pc$^{-3}$. The abundance of these ultradense haloes substantially alters the predictions of many PBH scenarios and enables a new range of observational tests.

\section{Formation of ultradense haloes}\label{sec:collapse}

During the radiation-dominated epoch, primordial curvature perturbations $\zeta$ at the scale wavenumber $k$ cause linear-order dark matter density perturbations $\delta$ to grow as
\begin{equation}\label{D}
    \delta(k,a) = I_1 \zeta(k) \log(I_2 a/a_H) = I_1 \zeta(k) \log\left(\sqrt{2}I_2\frac{k}{\keq}\frac{a}{\aeq}\right) 
\end{equation}
when $a\gg a_H$, where $a_H = 2^{-1/2} (\keq/k) \aeq$ is the expansion factor at horizon entry, $I_1\simeq 6.4$, and $I_2\simeq 0.47$ \citep{1996ApJ...471..542H}. Here $\aeq\simeq 3\times 10^{-4}$ and $\keq\simeq 0.01$~Mpc$^{-1}$ are the expansion factor and horizon scale at matter-radiation equality. The growth described by Eq.~(\ref{D}) arises from unaccelerated particle drift, in which particles cover comoving distances logarithmic in $a$. The particles are initially set in motion by the transient peculiar potential at horizon entry, before the radiation pressure homogenizes it. By adopting Eq.~(\ref{D}), we thus assume that the dark matter is nonrelativistic and decoupled from the radiation during this horizon-entry gravitational kick.

Since peculiar gravitational forces are negligible, an ellipsoidal collapse treatment is trivial because we can simply allow each axis to drift independently. In particular, if a region of scale $k^{-1}$ has an initial (transient) tidal field with ellipticity $e$ and prolateness $p$, then its density evolves as
\begin{equation}\label{density}
    \rho/\rhoM = |1-\lambda_1\delta(k,a)|^{-1}|1-\lambda_2\delta(k,a)|^{-1}|1-\lambda_3\delta(k,a)|^{-1},
\end{equation}
where $\rhoM$ is the average dark matter density and $\lambda_1=(1+3e+p)/3$, $\lambda_2=(1-2p)/3$, and $\lambda_3=(1-3e+p)/3$. The last axis to collapse does so at the scale factor $\ac$ when $\lambda_3\delta(k,\ac) = 1$, i.e. at the critical linear density contrast
\begin{equation}\label{dcep}
    \dc = 3/(1-3e+p).
\end{equation}

For a density contrast $\delta$ in a Gaussian random density field with rms contrast $\sigma$, the most probable values of $e$ and $p$ are $e=(\sqrt{5}\delta/\sigma)^{-1}$ and $p=0$, respectively \citep{2001MNRAS.323....1S}. For example, a typical $3\sigma$ density peak has a collapse threshold of about $\dc\simeq 5$. For such a peak to collapse by $a\simeq 10a_H$ requires $\zeta\gtrsim 0.5$. However, only $\zeta\gtrsim 0.15$ is required to induce collapse of such a peak by $a\simeq 300a_H$. In contrast, $\zeta\sim 1$ is required to form a primordial black hole \citep{2004PhRvD..70d1502G}. Thus, it is already clear that minihaloes can potentially far outnumber primordial black holes.

However, the formation of haloes from collapsed structures cannot be taken for granted during the radiation epoch. \citet{2019PhRvD.100j3010B} demonstrated in a numerical simulation that the collapse of a peak yields a virialized halo only when the collapsed region becomes locally matter dominated. Consequently, it is not guaranteed that a halo will form at the moment of collapse. On the other hand, \citet{2019PhRvD.100j3010B} also noted that collapsed regions can be orders of magnitude denser than the dark matter average. In particular, we estimate in Appendix~\ref{sec:collapsed} that the density of a collapsed region is of order \chg{$e^{-2}\rhoM$,} where $e$ is the ellipticity of the initial tidal field again. For the typical ellipticity $e\simeq 0.15$ of the tidal field at a $3\sigma$ peak, local matter domination thus occurs at $a\sim e^2\aeq\simeq 0.02\aeq$, leading to halo formation long before the matter-dominated epoch begins. Being proportional to the mean cosmic density at their formation time, the density inside these haloes would be extraordinarily high.

If we replace $e$ and $p$ in Eq.~(\ref{dcep}) by their their most probable values, $e=(\sqrt{5}\dc/\sigma)^{-1}$ and $p=0$, then we obtain the collapse threshold
$\dc = 3(1 + \sigma/\sqrt{5})$
as a function of $\sigma$. In the excursion set formalism \citep{1991ApJ...379..440B}, this threshold corresponds to the moving barrier
\begin{equation}
    B(S)\equiv 3(1+\sqrt{S/5}), 
\end{equation}
where $S\equiv \sigma^2$. For a Gaussian random walk, the distribution of first barrier crossings in this scenario is well approximated by
\begin{align}\label{crossing}
    F(S) &= \frac{3 + 0.556\sqrt{S}}{\sqrt{2\pi S^3}}
    \exp\left[-\frac{B(S)^2}{2S}\right] \left(1+\frac{S}{400}\right)^{-0.15},
\end{align}
which we verified by Monte Carlo simulation. This distribution leads to the Press-Schechter halo mass function
\begin{equation}\label{massfunction}
    \frac{\diff n}{\diff\log M}=\sqrt{\frac{2}{\pi}}
    \frac{(\nu+0.556)\e^{-\frac{1}{2}(\nu+1.34)^2}}{\left(1+0.0225\nu^{-2}\right)^{0.15}}
    \frac{\diff\log\nu}{\diff\log M} \frac{\rhoMo}{M},
\end{equation}
where $\rhoMo\simeq 33$~M$_\odot$\,kpc$^{-3}$ is the comoving dark matter density and $\nu\equiv 3/\sigma_M$. Here, $\sigma_M$ is the rms density contrast smoothed on the mass scale $M$, which we evaluate using a sharp-$k$ filter in order to accommodate power spectra that deviate significantly from scale invariance \citep{2013MNRAS.428.1774B}. That is,
\begin{equation}
    \sigma_M^2 = \int_0^{k_M}\frac{\diff k}{k}\mathcal{P}(k,a)
\end{equation}
with $M\equiv 6\pi^2\rhoMo k_M^{-3}$ \citep{1993MNRAS.262..627L}, which implies also
\begin{equation}
    \frac{\diff\log\nu}{\diff\log M} = \frac{\mathcal{P}(k_M,a)}{6\sigma_M^2}.
\end{equation}
Here, $\mathcal{P}(k,a)\equiv [k^3/(2\pi^2)]P(k,a)$ is the dimensionless matter power spectrum, which is a function of time. In particular, Eq.~(\ref{D}) implies that
\begin{equation}
    \mathcal{P}(k,a) = I_1^2 \left[\log\left(\sqrt{2}I_2\frac{k}{\keq}\frac{a}{\aeq}\right)\right]^2 \mathcal{P}_\zeta(k), 
\end{equation}
where $\mathcal{P}_\zeta(k)$ is the dimensionless power spectrum of primordial curvature perturbations.

Note that while we focused here on structure arising from curvature perturbations, halo formation from isocurvature perturbations during the radiation epoch \citep[e.g.][]{1994PhRvD..50..769K} is even more favourable in comparison to PBH formation. For adiabatic curvature perturbations, haloes are disadvantaged in relation to PBHs by the requirement that local matter domination be achieved before peculiar gravitational forces become significant. PBHs, in contrast, can form from radiation perturbations alone. But in the case of structure arising from isocurvature perturbations, both the PBHs and the haloes form essentially out of matter, so the haloes are no longer so disadvantaged. Similar considerations apply to PBHs that form during an early matter-dominated phase \citep{2019EPJC...79..246B}.

\section{Ultradense haloes and PBHs}

Due to the arguments above, ultradense minihaloes are generally expected to vastly outnumber PBHs in scenarios where there is nonrelativistic dark matter decoupled from the radiation at the time that the large-amplitude initial density variations are entering the horizon. This dark matter could be either particle dark matter or smaller PBHs, implying two possibilities:
\begin{enumerate}
    \item PBHs are only a fraction of the dark matter, and the ultradense haloes form out of particle dark matter.
    \item PBHs are all of the dark matter but have a mass function that extends over many orders of magnitude,\footnote{The PBH mass spectrum clearly must extend over multiple orders of magnitude in order for an ultradense halo to contain a significant number of smaller PBHs, but the question of precisely how broad it must be is subtle. A less extended mass spectrum leads to ultradense haloes consisting of fewer PBHs, which could cause collisional relaxation effects to more strongly affect halo structures \citep[e.g.][]{2011PhRvD..84l4031C}. We leave this question for future work.} so that the ultradense haloes are clusters of much smaller PBHs.
\end{enumerate}
In the latter case, the PBH mass spectrum must also be weighted towards low masses, but such weighting is naturally expected to arise for broad primordial power spectra \citep{2020PhLB..80735550D}. We now discuss two examples of this case.

\begin{figure}
	\centering
	\includegraphics[width=\columnwidth]{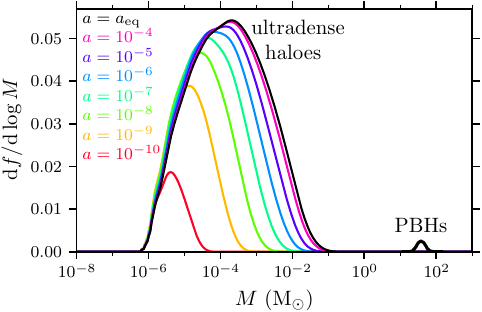}
	\caption{Differential dark matter mass fraction in ultradense haloes (left) and PBHs (right) in the double inflation scenario of \citet{2018PhRvD..97d3514I}. We consider regions that collapse to form haloes by matter-radiation equality (black) as well as the subset of these regions that collapse by earlier times (colours); note that collapse precedes halo formation but does not necessarily induce it immediately (see the text). In total, these ultradense haloes in the range $10^{-6}$ to $10^{-1}$~M$_\odot$ \chg{contain} roughly 40 per cent of the dark matter. In contrast, $10$ to $100$~M$_\odot$ PBHs, which form from fluctuations at the same scale, comprise only about 0.2 per cent of the dark matter.}
	\label{fig:mass_function}
\end{figure}

As our first example, let us consider the scenario presented by \citet{2018PhRvD..97d3514I}. Here, a double inflation model yields a complicated primordial power spectrum that produces asteroid-mass PBHs that comprise almost all of the dark matter while also producing a small abundance of $10$ to $100$~M$_\odot$ PBHs to explain LVK binary coalescence detections \citep[e.g.][]{2016PhRvX...6d1015A}. We then expect that the density variations that produce the $10$ to $100$~M$_\odot$ PBHs also create ultradense minihaloes consisting of asteroid-mass PBHs. In particular, Fig.~\ref{fig:mass_function} shows the differential mass fraction $\diff f/\diff\log M = (M/\rhoMo)\diff n/\diff\log M$ in ultradense haloes that form during the radiation epoch; note that we include only haloes that form from the largest-scale spike in the primordial power spectrum \citep[which is the same for both models presented in][]{2018PhRvD..97d3514I}. In total, $10$ to $100$~M$_\odot$ PBHs comprise only about 0.2 per cent of the dark matter, but the boosted density variations necessary to produce them cause about 40 per cent of the dark matter mass to reside in ultradense haloes between earth mass and solar mass. Note that for adiabatic initial fluctuations, individual ultradense haloes are generically expected to be somewhat less massive than PBHs arising at the same length scale because haloes form from matter, which is $a/\aeq$ times as abundant as the radiation that produces PBHs.

\begin{figure}
	\centering
	\includegraphics[width=\columnwidth]{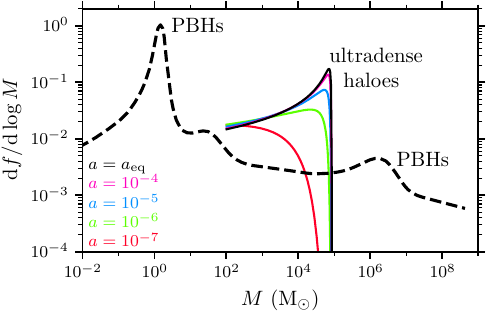}
	\caption{Differential dark matter mass fraction in ultradense haloes (solid curves) and PBHs (dashed curve) in the PBH scenario of \citet{2021PDU....3100755C}. We consider regions that collapse to form haloes by matter-radiation equality (black) as well as the subset of these regions that collapse by earlier times (colours). The tail of PBHs larger than $10^3$~M$_\odot$ act as supermassive black hole seeds, but this tail is accompanied by an abundance of ultradense haloes that form out of much smaller $\mathcal{O}(1)$~M$_\odot$ PBHs. About \chg{25} per cent of the dark matter resides in ultradense haloes larger than $10^3$~M$_\odot$.}
	\label{fig:mass_function_b}
\end{figure}

We next consider the scenario of \citet{2021PDU....3100755C}. Here, the primordial power spectrum is
\begin{equation}\label{pkflat}
    \mathcal{P}_\zeta(k) =
    \begin{cases}
        \As\left[k/(0.05~\mathrm{Mpc}^{-1})\right]^{\ns-1} & \text{if}\ \ k < k_1 \\
        A_1\left[k/(10^6~\mathrm{Mpc}^{-1})\right]^{\ns-1} & \text{if}\ \ k > k_1,
    \end{cases}
\end{equation}
where $\As=2.1\times 10^{-9}$ and $\ns=0.96$ to match data from the cosmic microwave background \citep{2020A&A...641A...6P}, but the power is boosted to $A_1\simeq 0.022$ at small scales \chg{$k>k_1\simeq 190$~Mpc$^{-1}$} in order to produce PBHs in the right abundance to comprise all of the dark matter. This spectrum is featureless and nearly scale invariant for $k>k_1$, but it yields the nontrivial PBH mass function shown in Fig.~\ref{fig:mass_function_b} due to the nontrivial thermal history of the early universe. In this scenario, the bulk of the dark matter consists of $\mathcal{O}(1)$~M$_\odot$ PBHs, but a tail of much larger PBHs act as seeds for the supermassive black holes found at the centres of galaxies. However, density variations on scales large enough to produce such PBHs also cause the solar mass PBHs to cluster into ultradense haloes up to \chg{nearly $10^5$~M$_\odot$}.\footnote{\chg{The sharp cutoff near $10^5$~M$_\odot$ is a consequence of the sharp cutoff in power in Eq.~(\ref{pkflat}). The scenario in \cite{2021PDU....3100755C} is ad hoc, not drawn from an inflationary model, and such abrupt features are unlikely in a more realistic scenario \citep{2019JCAP...06..028B}.}} Whereas PBHs larger than $10^3$~M$_\odot$ comprise only 3 per cent of the dark matter, 30 per cent of the dark matter resides in ultradense haloes above this mass scale.

\chg{More generally, if haloes of mass $M_\mathrm{halo}$ and PBHs of mass $M_\mathrm{PBH}$ form from density fluctuations of the same scale, then
\begin{equation}\label{masses}
    M_\mathrm{halo}\sim M_\mathrm{PBH}^{3/2} \Meq^{-1/2}
\end{equation}
approximately relates the two mass scales, where $\Meq\simeq 3\times 10^{17}$~M$_\odot$ is the horizon mass at matter-radiation equality. As we discussed above, the difference between the mass scales arises because PBHs form from (predominantly) radiation while haloes form from matter. Equation~(\ref{masses}) can be viewed as a lower limit on the mass scale for ultradense haloes, because larger-scale density variations may also be present that are extreme enough to produce ultradense haloes but not extreme enough to make PBHs. This is the case in Fig.~\ref{fig:mass_function}, for example, as 10 to 100~M$_\odot$ PBHs correspond to haloes around the $10^{-6}$~M$_\odot$ scale according to Eq.~(\ref{masses}), and yet ultradense haloes as large as $10^{-2}$~M$_\odot$ are predicted.}

In Figs. \ref{fig:mass_function} and~\ref{fig:mass_function_b}, we show not only the halo distribution at matter-radiation equality (in black) but also the distribution of collapsed regions at earlier times (colours). As we discussed in Section~\ref{sec:collapse}, a collapsed region forms a halo once it becomes locally matter dominated, which occurs at a time of about $a\sim e^2\aeq$, where $e$ is the ellipticity of the region's initial tidal field. For typical ellipticity values around $0.1$ to $0.3$, local matter domination occurs around $a\sim 10^{-5}$ within regions that have collapsed by then. Since Figs. \ref{fig:mass_function} and~\ref{fig:mass_function_b} show that the distribution of collapsed regions at $a=10^{-5}$ is comparable to that at $a=\aeq$, we conclude that most ultradense haloes have formation times $\af\sim 10^{-5}$.

Meanwhile, a halo's internal density is of order $10^3$ times the mean density of the universe at its formation time $\af$ \citep{2023MNRAS.518.3509D}, i.e.
\begin{equation}
    \rho_\mathrm{halo} \sim 10^3 (1+\aeq/\af) \af^{-3} \rhoMo,
\end{equation}
where $\rhoMo\simeq 33$~M$_\odot$\,kpc$^{-3}$ is again the comoving dark matter density. The formation time $\af\sim 10^{-5}$ for ultradense haloes thus implies that they have internal densities $\rho_\mathrm{halo}\sim 10^{12}$~M$_\odot$\,pc$^{-3}$. These ultradense haloes are about $10^{12}$ times denser than the first haloes that would form around redshift 30 in a standard cold dark matter cosmology \citep[e.g.][]{2022arXiv220911237D} and of order $10^{19}$ times denser than the cosmological mean today.

\chg{Finally, we emphasize that ultradense haloes are broadly expected to persist through later evolution. We first comment that a halo's deep internal structure is largely unaffected by material accreted later \citep[e.g.][]{2023MNRAS.518.3509D}, which is why the halo properties that we study during the radiation epoch remain relevant at later times. Also, the extreme internal density of ultradense haloes makes them highly resistant to disruptive effects, such as tidal stripping \citep[e.g.][]{2022arXiv220700604S}, inside larger objects like galactic haloes. Mergers between ultradense haloes are of concern, as they could deplete the number of these haloes. However, while further study is needed to fully understand the impact of mergers, their impact is likely to be modest. The amplitudes of initial density variations must drop steeply at larger mass scales, which would make halo mergers rare. This is reflected, for example, in how slowly the halo mass distribution in Fig.~\ref{fig:mass_function} shifts to higher masses once the ultradense halo population is established around $a\sim 10^{-5}$. We also remark that irrespective of their frequency, mergers do not deplete the mass fraction in haloes, and they typically do not reduce the haloes' characteristic internal density values either \citep{2019MNRAS.487.1008D,2019PhRvD.100b3523D}.}


\section{Conclusion}\label{sec:conclusion}

If PBHs have a mass function that extends over many orders of magnitude, or if they comprise only a fraction of the dark matter, then they are expected to be accompanied by ultradense minihaloes with internal densities of order $10^{12}$ times that of standard cold dark matter haloes. These ultradense haloes form during the radiation-dominated epoch from large-amplitude initial density variations similar to those that are necessary to produce PBHs. However, the haloes \chg{contain} an $\mathcal{O}(1)$ fraction of the dark matter even when PBHs arising at the same length scales contribute only at the per cent level.

For PBHs in the 10 to 100~M$_\odot$ mass range, which have been suggested to explain LVK observations, ultradense haloes are produced in the earth-mass to solar-mass range. Such minihaloes can be detected by pulsar timing distortions \citep{2019PhRvD.100b3003D} or gravitational lensing \citep{2011ApJ...729...49E,2020PhRvD.101h3013C,2020AJ....159...49D,2022arXiv220805957O,2022arXiv221002078C}. \chg{Indeed, limits from microlensing on the abundance of compact objects in this mass range may place new constraints on the possibility that LVK events are of primordial origin \citep{https://doi.org/10.48550/arxiv.2301.13171}.}

Much larger PBHs, which could act as seeds for supermassive black holes, are accompanied by ultradense haloes potentially \chg{up to of order $10^5$~M$_\odot$}. These haloes are sufficiently massive that they could produce dynamical signatures within galaxies \citep{2022MNRAS.513.3682D}, affect early galaxy formation \citep{2022arXiv220714735I}, and alter the structures of galactic haloes via their gravitational scattering \citep{2022ApJ...929L..24L}. \chg{Limits imposed by stellar dynamics on the abundance of compact objects in this regime are severe \citep[e.g.][]{2016ApJ...824L..31B,2017PhRvL.119d1102K} and may rule out ultradense haloes significantly above a solar mass. We remark, however, that limits on spectral distortions in the cosmic microwave background already challenge such scenarios \citep{2012ApJ...758...76C}.}

\chg{There are also consequences related to the interiors of such dense systems.} In mixed dark matter scenarios with PBHs and \chg{thermally produced particle dark matter}, ultradense haloes would enormously boost the dark matter annihilation rate, potentially leading to detectable gamma rays \citep{2010PhRvD..82h3527J,2012PhRvD..85l5027B} or other signatures \citep[e.g.][]{2017JCAP...05..048C,2018PhRvL.121w1105S}, although these mixed scenarios are already tightly constrained \citep[e.g.][]{2010ApJ...720L..67L,2019PhRvD.100b3506A}. For broad PBH mass spectra, the presence of ultradense haloes implies that \chg{the PBHs that they contain} quickly become highly clustered, even if they are not clustered at formation \citep[][]{2019JCAP...11..001M}. Such clustering can significantly impact the binary coalescence rate \citep{2019PhRvD..99f3532B,2019JCAP...02..018R,2020JCAP...11..028D}. \chg{In these ways}, the ultradense haloes that accompany PBH formation enable an array of new tests of PBH scenarios.

\section*{Acknowledgements}

We thank Fabian Schmidt and Eiichiro Komatsu for helpful comments on the manuscript.

\section*{Data Availability}
 
No datasets were created or analysed during this study.



\bibliographystyle{mnras}
\bibliography{main}

\begin{thebibliography}{}
\makeatletter
\relax
\def\mn@urlcharsother{\let\do\@makeother \do\$\do\&\do\#\do\^\do\_\do\%\do\~}
\def\mn@doi{\begingroup\mn@urlcharsother \@ifnextchar [ {\mn@doi@}
  {\mn@doi@[]}}
\def\mn@doi@[#1]#2{\def\@tempa{#1}\ifx\@tempa\@empty \href
  {http://dx.doi.org/#2} {doi:#2}\else \href {http://dx.doi.org/#2} {#1}\fi
  \endgroup}
\def\mn@eprint#1#2{\mn@eprint@#1:#2::\@nil}
\def\mn@eprint@arXiv#1{\href {http://arxiv.org/abs/#1} {{\tt arXiv:#1}}}
\def\mn@eprint@dblp#1{\href {http://dblp.uni-trier.de/rec/bibtex/#1.xml}
  {dblp:#1}}
\def\mn@eprint@#1:#2:#3:#4\@nil{\def\@tempa {#1}\def\@tempb {#2}\def\@tempc
  {#3}\ifx \@tempc \@empty \let \@tempc \@tempb \let \@tempb \@tempa \fi \ifx
  \@tempb \@empty \def\@tempb {arXiv}\fi \@ifundefined
  {mn@eprint@\@tempb}{\@tempb:\@tempc}{\expandafter \expandafter \csname
  mn@eprint@\@tempb\endcsname \expandafter{\@tempc}}}

\bibitem[\protect\citeauthoryear{{Abbott} et~al.,}{{Abbott}
  et~al.}{2016}]{2016PhRvX...6d1015A}
{Abbott} B.~P.,  et~al., 2016, \mn@doi [Physical Review X]
  {10.1103/PhysRevX.6.041015}, \href
  {https://ui.adsabs.harvard.edu/abs/2016PhRvX...6d1015A} {6, 041015}

\bibitem[\protect\citeauthoryear{{Abbott} et~al.,}{{Abbott}
  et~al.}{2020}]{2020PhRvL.125j1102A}
{Abbott} R.,  et~al., 2020, \mn@doi [\prl] {10.1103/PhysRevLett.125.101102},
  \href {https://ui.adsabs.harvard.edu/abs/2020PhRvL.125j1102A} {125, 101102}

\bibitem[\protect\citeauthoryear{{Adamek}, {Byrnes}, {Gosenca}  \&
  {Hotchkiss}}{{Adamek} et~al.}{2019}]{2019PhRvD.100b3506A}
{Adamek} J.,  {Byrnes} C.~T.,  {Gosenca} M.,   {Hotchkiss} S.,  2019, \mn@doi
  [\prd] {10.1103/PhysRevD.100.023506}, \href
  {https://ui.adsabs.harvard.edu/abs/2019PhRvD.100b3506A} {100, 023506}

\bibitem[\protect\citeauthoryear{{Belotsky} et~al.,}{{Belotsky}
  et~al.}{2019}]{2019EPJC...79..246B}
{Belotsky} K.~M.,  et~al., 2019, \mn@doi [European Physical Journal C]
  {10.1140/epjc/s10052-019-6741-4}, \href
  {https://ui.adsabs.harvard.edu/abs/2019EPJC...79..246B} {79, 246}

\bibitem[\protect\citeauthoryear{{Benson} et~al.,}{{Benson}
  et~al.}{2013}]{2013MNRAS.428.1774B}
{Benson} A.~J.,  et~al., 2013, \mn@doi [\mnras] {10.1093/mnras/sts159}, \href
  {https://ui.adsabs.harvard.edu/abs/2013MNRAS.428.1774B} {428, 1774}

\bibitem[\protect\citeauthoryear{{Berezinsky}, {Dokuchaev}, {Eroshenko},
  {Kachelrie{\ss}}  \& {Solberg}}{{Berezinsky}
  et~al.}{2010}]{2010PhRvD..81j3529B}
{Berezinsky} V.,  {Dokuchaev} V.,  {Eroshenko} Y.,  {Kachelrie{\ss}} M.,
  {Solberg} M.~A.,  2010, \mn@doi [\prd] {10.1103/PhysRevD.81.103529}, \href
  {https://ui.adsabs.harvard.edu/abs/2010PhRvD..81j3529B} {81, 103529}

\bibitem[\protect\citeauthoryear{{Berezinsky}, {Dokuchaev}  \&
  {Eroshenko}}{{Berezinsky} et~al.}{2013}]{2013JCAP...11..059B}
{Berezinsky} V.~S.,  {Dokuchaev} V.~I.,   {Eroshenko} Y.~N.,  2013, \mn@doi
  [\jcap] {10.1088/1475-7516/2013/11/059}, \href
  {https://ui.adsabs.harvard.edu/abs/2013JCAP...11..059B} {2013, 059}

\bibitem[\protect\citeauthoryear{{Bird}, {Cholis}, {Mu{\~n}oz},
  {Ali-Ha{\"\i}moud}, {Kamionkowski}, {Kovetz}, {Raccanelli}  \&
  {Riess}}{{Bird} et~al.}{2016}]{2016PhRvL.116t1301B}
{Bird} S.,  {Cholis} I.,  {Mu{\~n}oz} J.~B.,  {Ali-Ha{\"\i}moud} Y.,
  {Kamionkowski} M.,  {Kovetz} E.~D.,  {Raccanelli} A.,   {Riess} A.~G.,  2016,
  \mn@doi [\prl] {10.1103/PhysRevLett.116.201301}, \href
  {https://ui.adsabs.harvard.edu/abs/2016PhRvL.116t1301B} {116, 201301}

\bibitem[\protect\citeauthoryear{{Blanco}, {Delos}, {Erickcek}  \&
  {Hooper}}{{Blanco} et~al.}{2019}]{2019PhRvD.100j3010B}
{Blanco} C.,  {Delos} M.~S.,  {Erickcek} A.~L.,   {Hooper} D.,  2019, \mn@doi
  [\prd] {10.1103/PhysRevD.100.103010}, \href
  {https://ui.adsabs.harvard.edu/abs/2019PhRvD.100j3010B} {100, 103010}

\bibitem[\protect\citeauthoryear{{Boldrini}, {Miki}, {Wagner}, {Mohayaee},
  {Silk}  \& {Arbey}}{{Boldrini} et~al.}{2020}]{2020MNRAS.492.5218B}
{Boldrini} P.,  {Miki} Y.,  {Wagner} A.~Y.,  {Mohayaee} R.,  {Silk} J.,
  {Arbey} A.,  2020, \mn@doi [\mnras] {10.1093/mnras/staa150}, \href
  {https://ui.adsabs.harvard.edu/abs/2020MNRAS.492.5218B} {492, 5218}

\bibitem[\protect\citeauthoryear{{Bond}, {Cole}, {Efstathiou}  \&
  {Kaiser}}{{Bond} et~al.}{1991}]{1991ApJ...379..440B}
{Bond} J.~R.,  {Cole} S.,  {Efstathiou} G.,   {Kaiser} N.,  1991, \mn@doi
  [\apj] {10.1086/170520}, \href
  {https://ui.adsabs.harvard.edu/abs/1991ApJ...379..440B} {379, 440}

\bibitem[\protect\citeauthoryear{{Brandt}}{{Brandt}}{2016}]{2016ApJ...824L..31B}
{Brandt} T.~D.,  2016, \mn@doi [\apjl] {10.3847/2041-8205/824/2/L31}, \href
  {https://ui.adsabs.harvard.edu/abs/2016ApJ...824L..31B} {824, L31}

\bibitem[\protect\citeauthoryear{{Bringmann}, {Scott}  \& {Akrami}}{{Bringmann}
  et~al.}{2012}]{2012PhRvD..85l5027B}
{Bringmann} T.,  {Scott} P.,   {Akrami} Y.,  2012, \mn@doi [\prd]
  {10.1103/PhysRevD.85.125027}, \href
  {https://ui.adsabs.harvard.edu/abs/2012PhRvD..85l5027B} {85, 125027}

\bibitem[\protect\citeauthoryear{{Bringmann}, {Depta}, {Domcke}  \&
  {Schmidt-Hoberg}}{{Bringmann} et~al.}{2019}]{2019PhRvD..99f3532B}
{Bringmann} T.,  {Depta} P.~F.,  {Domcke} V.,   {Schmidt-Hoberg} K.,  2019,
  \mn@doi [\prd] {10.1103/PhysRevD.99.063532}, \href
  {https://ui.adsabs.harvard.edu/abs/2019PhRvD..99f3532B} {99, 063532}

\bibitem[\protect\citeauthoryear{{Byrnes}, {Hindmarsh}, {Young}  \&
  {Hawkins}}{{Byrnes} et~al.}{2018}]{2018JCAP...08..041B}
{Byrnes} C.~T.,  {Hindmarsh} M.,  {Young} S.,   {Hawkins} M. R.~S.,  2018,
  \mn@doi [\jcap] {10.1088/1475-7516/2018/08/041}, \href
  {https://ui.adsabs.harvard.edu/abs/2018JCAP...08..041B} {2018, 041}

\bibitem[\protect\citeauthoryear{{Byrnes}, {Cole}  \& {Patil}}{{Byrnes}
  et~al.}{2019}]{2019JCAP...06..028B}
{Byrnes} C.~T.,  {Cole} P.~S.,   {Patil} S.~P.,  2019, \mn@doi [\jcap]
  {10.1088/1475-7516/2019/06/028}, \href
  {https://ui.adsabs.harvard.edu/abs/2019JCAP...06..028B} {2019, 028}

\bibitem[\protect\citeauthoryear{{Cai}, {Chen}, {Wang}  \& {Yang}}{{Cai}
  et~al.}{2022}]{2022arXiv221002078C}
{Cai} R.-G.,  {Chen} T.,  {Wang} S.-J.,   {Yang} X.-Y.,  2022, arXiv e-prints,
  \href {https://ui.adsabs.harvard.edu/abs/2022arXiv221002078C} {p.
  arXiv:2210.02078}

\bibitem[\protect\citeauthoryear{{Carr} \& {K{\"u}hnel}}{{Carr} \&
  {K{\"u}hnel}}{2020}]{2020ARNPS..70..355C}
{Carr} B.,  {K{\"u}hnel} F.,  2020, \mn@doi [Annual Review of Nuclear and
  Particle Science] {10.1146/annurev-nucl-050520-125911}, \href
  {https://ui.adsabs.harvard.edu/abs/2020ARNPS..70..355C} {70, 355}

\bibitem[\protect\citeauthoryear{{Carr} \& {Silk}}{{Carr} \&
  {Silk}}{2018}]{2018MNRAS.478.3756C}
{Carr} B.,  {Silk} J.,  2018, \mn@doi [\mnras] {10.1093/mnras/sty1204}, \href
  {https://ui.adsabs.harvard.edu/abs/2018MNRAS.478.3756C} {478, 3756}

\bibitem[\protect\citeauthoryear{{Carr}, {Clesse}, {Garc{\'\i}a-Bellido}  \&
  {K{\"u}hnel}}{{Carr} et~al.}{2021a}]{2021PDU....3100755C}
{Carr} B.,  {Clesse} S.,  {Garc{\'\i}a-Bellido} J.,   {K{\"u}hnel} F.,  2021a,
  \mn@doi [Physics of the Dark Universe] {10.1016/j.dark.2020.100755}, \href
  {https://ui.adsabs.harvard.edu/abs/2021PDU....3100755C} {31, 100755}

\bibitem[\protect\citeauthoryear{{Carr}, {Kohri}, {Sendouda}  \&
  {Yokoyama}}{{Carr} et~al.}{2021b}]{2021RPPh...84k6902C}
{Carr} B.,  {Kohri} K.,  {Sendouda} Y.,   {Yokoyama} J.,  2021b, \mn@doi
  [Reports on Progress in Physics] {10.1088/1361-6633/ac1e31}, \href
  {https://ui.adsabs.harvard.edu/abs/2021RPPh...84k6902C} {84, 116902}

\bibitem[\protect\citeauthoryear{{Chisholm}}{{Chisholm}}{2011}]{2011PhRvD..84l4031C}
{Chisholm} J.~R.,  2011, \mn@doi [\prd] {10.1103/PhysRevD.84.124031}, \href
  {https://ui.adsabs.harvard.edu/abs/2011PhRvD..84l4031C} {84, 124031}

\bibitem[\protect\citeauthoryear{{Chluba}, {Erickcek}  \& {Ben-Dayan}}{{Chluba}
  et~al.}{2012}]{2012ApJ...758...76C}
{Chluba} J.,  {Erickcek} A.~L.,   {Ben-Dayan} I.,  2012, \mn@doi [\apj]
  {10.1088/0004-637X/758/2/76}, \href
  {https://ui.adsabs.harvard.edu/abs/2012ApJ...758...76C} {758, 76}

\bibitem[\protect\citeauthoryear{{Clark}, {Iwanus}, {Elahi}, {Lewis}  \&
  {Scott}}{{Clark} et~al.}{2017}]{2017JCAP...05..048C}
{Clark} H.~A.,  {Iwanus} N.,  {Elahi} P.~J.,  {Lewis} G.~F.,   {Scott} P.,
  2017, \mn@doi [\jcap] {10.1088/1475-7516/2017/05/048}, \href
  {https://ui.adsabs.harvard.edu/abs/2017JCAP...05..048C} {2017, 048}

\bibitem[\protect\citeauthoryear{{Croon}, {McKeen}  \& {Raj}}{{Croon}
  et~al.}{2020}]{2020PhRvD.101h3013C}
{Croon} D.,  {McKeen} D.,   {Raj} N.,  2020, \mn@doi [\prd]
  {10.1103/PhysRevD.101.083013}, \href
  {https://ui.adsabs.harvard.edu/abs/2020PhRvD.101h3013C} {101, 083013}

\bibitem[\protect\citeauthoryear{{Dai} \& {Miralda-Escud{\'e}}}{{Dai} \&
  {Miralda-Escud{\'e}}}{2020}]{2020AJ....159...49D}
{Dai} L.,  {Miralda-Escud{\'e}} J.,  2020, \mn@doi [\aj]
  {10.3847/1538-3881/ab5e83}, \href
  {https://ui.adsabs.harvard.edu/abs/2020AJ....159...49D} {159, 49}

\bibitem[\protect\citeauthoryear{{De Luca}, {Franciolini}  \& {Riotto}}{{De
  Luca} et~al.}{2020a}]{2020PhLB..80735550D}
{De Luca} V.,  {Franciolini} G.,   {Riotto} A.,  2020a, \mn@doi [Physics
  Letters B] {10.1016/j.physletb.2020.135550}, \href
  {https://ui.adsabs.harvard.edu/abs/2020PhLB..80735550D} {807, 135550}

\bibitem[\protect\citeauthoryear{{De Luca}, {Desjacques}, {Franciolini}  \&
  {Riotto}}{{De Luca} et~al.}{2020b}]{2020JCAP...11..028D}
{De Luca} V.,  {Desjacques} V.,  {Franciolini} G.,   {Riotto} A.,  2020b,
  \mn@doi [\jcap] {10.1088/1475-7516/2020/11/028}, \href
  {https://ui.adsabs.harvard.edu/abs/2020JCAP...11..028D} {2020, 028}

\bibitem[\protect\citeauthoryear{Delos \& Franciolini}{Delos \&
  Franciolini}{2023}]{https://doi.org/10.48550/arxiv.2301.13171}
Delos M.~S.,  Franciolini G.,  2023, \mn@doi [arXiv e-prints]
  {10.48550/ARXIV.2301.13171}, \href
  {https://ui.adsabs.harvard.edu/abs/arXiv:2301.13171} {p. arXiv:2301.13171}

\bibitem[\protect\citeauthoryear{{Delos} \& {Schmidt}}{{Delos} \&
  {Schmidt}}{2022}]{2022MNRAS.513.3682D}
{Delos} M.~S.,  {Schmidt} F.,  2022, \mn@doi [\mnras] {10.1093/mnras/stac1022},
  \href {https://ui.adsabs.harvard.edu/abs/2022MNRAS.513.3682D} {513, 3682}

\bibitem[\protect\citeauthoryear{{Delos} \& {White}}{{Delos} \&
  {White}}{2022}]{2022arXiv220911237D}
{Delos} M.~S.,  {White} S. D.~M.,  2022, arXiv e-prints, \href
  {https://ui.adsabs.harvard.edu/abs/2022arXiv220911237D} {p. arXiv:2209.11237}

\bibitem[\protect\citeauthoryear{{Delos} \& {White}}{{Delos} \&
  {White}}{2023}]{2023MNRAS.518.3509D}
{Delos} M.~S.,  {White} S. D.~M.,  2023, \mn@doi [\mnras]
  {10.1093/mnras/stac3373}, \href
  {https://ui.adsabs.harvard.edu/abs/2023MNRAS.518.3509D} {518, 3509}

\bibitem[\protect\citeauthoryear{{Delos}, {Erickcek}, {Bailey}  \&
  {Alvarez}}{{Delos} et~al.}{2018a}]{2018PhRvD..97d1303D}
{Delos} M.~S.,  {Erickcek} A.~L.,  {Bailey} A.~P.,   {Alvarez} M.~A.,  2018a,
  \mn@doi [\prd] {10.1103/PhysRevD.97.041303}, \href
  {https://ui.adsabs.harvard.edu/abs/2018PhRvD..97d1303D} {97, 041303}

\bibitem[\protect\citeauthoryear{{Delos}, {Erickcek}, {Bailey}  \&
  {Alvarez}}{{Delos} et~al.}{2018b}]{2018PhRvD..98f3527D}
{Delos} M.~S.,  {Erickcek} A.~L.,  {Bailey} A.~P.,   {Alvarez} M.~A.,  2018b,
  \mn@doi [\prd] {10.1103/PhysRevD.98.063527}, \href
  {https://ui.adsabs.harvard.edu/abs/2018PhRvD..98f3527D} {98, 063527}

\bibitem[\protect\citeauthoryear{{Delos}, {Bruff}  \& {Erickcek}}{{Delos}
  et~al.}{2019}]{2019PhRvD.100b3523D}
{Delos} M.~S.,  {Bruff} M.,   {Erickcek} A.~L.,  2019, \mn@doi [\prd]
  {10.1103/PhysRevD.100.023523}, \href
  {https://ui.adsabs.harvard.edu/abs/2019PhRvD.100b3523D} {100, 023523}

\bibitem[\protect\citeauthoryear{{Dokuchaev} \& {Eroshenko}}{{Dokuchaev} \&
  {Eroshenko}}{2002}]{2002JETP...94....1D}
{Dokuchaev} V.~I.,  {Eroshenko} Y.~N.,  2002, \mn@doi [Soviet Journal of
  Experimental and Theoretical Physics] {10.1134/1.1448602}, \href
  {https://ui.adsabs.harvard.edu/abs/2002JETP...94....1D} {94, 1}

\bibitem[\protect\citeauthoryear{{Drakos}, {Taylor}, {Berrouet}, {Robotham}  \&
  {Power}}{{Drakos} et~al.}{2019}]{2019MNRAS.487.1008D}
{Drakos} N.~E.,  {Taylor} J.~E.,  {Berrouet} A.,  {Robotham} A. S.~G.,
  {Power} C.,  2019, \mn@doi [\mnras] {10.1093/mnras/stz1307}, \href
  {https://ui.adsabs.harvard.edu/abs/2019MNRAS.487.1008D} {487, 1008}

\bibitem[\protect\citeauthoryear{{Dror}, {Ramani}, {Trickle}  \&
  {Zurek}}{{Dror} et~al.}{2019}]{2019PhRvD.100b3003D}
{Dror} J.~A.,  {Ramani} H.,  {Trickle} T.,   {Zurek} K.~M.,  2019, \mn@doi
  [\prd] {10.1103/PhysRevD.100.023003}, \href
  {https://ui.adsabs.harvard.edu/abs/2019PhRvD.100b3003D} {100, 023003}

\bibitem[\protect\citeauthoryear{{Erickcek} \& {Law}}{{Erickcek} \&
  {Law}}{2011}]{2011ApJ...729...49E}
{Erickcek} A.~L.,  {Law} N.~M.,  2011, \mn@doi [\apj]
  {10.1088/0004-637X/729/1/49}, \href
  {https://ui.adsabs.harvard.edu/abs/2011ApJ...729...49E} {729, 49}

\bibitem[\protect\citeauthoryear{{Escriv{\`a}}, {Bagui}  \&
  {Clesse}}{{Escriv{\`a}} et~al.}{2022}]{2022arXiv220906196E}
{Escriv{\`a}} A.,  {Bagui} E.,   {Clesse} S.,  2022, arXiv e-prints, \href
  {https://ui.adsabs.harvard.edu/abs/2022arXiv220906196E} {p. arXiv:2209.06196}

\bibitem[\protect\citeauthoryear{{Franciolini} et~al.,}{{Franciolini}
  et~al.}{2022}]{2022PhRvD.105h3526F}
{Franciolini} G.,  et~al., 2022, \mn@doi [\prd] {10.1103/PhysRevD.105.083526},
  \href {https://ui.adsabs.harvard.edu/abs/2022PhRvD.105h3526F} {105, 083526}

\bibitem[\protect\citeauthoryear{{Gosenca}, {Adamek}, {Byrnes}  \&
  {Hotchkiss}}{{Gosenca} et~al.}{2017}]{2017PhRvD..96l3519G}
{Gosenca} M.,  {Adamek} J.,  {Byrnes} C.~T.,   {Hotchkiss} S.,  2017, \mn@doi
  [\prd] {10.1103/PhysRevD.96.123519}, \href
  {https://ui.adsabs.harvard.edu/abs/2017PhRvD..96l3519G} {96, 123519}

\bibitem[\protect\citeauthoryear{{Green} \& {Kavanagh}}{{Green} \&
  {Kavanagh}}{2021}]{2021JPhG...48d3001G}
{Green} A.~M.,  {Kavanagh} B.~J.,  2021, \mn@doi [Journal of Physics G Nuclear
  Physics] {10.1088/1361-6471/abc534}, \href
  {https://ui.adsabs.harvard.edu/abs/2021JPhG...48d3001G} {48, 043001}

\bibitem[\protect\citeauthoryear{{Green}, {Liddle}, {Malik}  \&
  {Sasaki}}{{Green} et~al.}{2004}]{2004PhRvD..70d1502G}
{Green} A.~M.,  {Liddle} A.~R.,  {Malik} K.~A.,   {Sasaki} M.,  2004, \mn@doi
  [\prd] {10.1103/PhysRevD.70.041502}, \href
  {https://ui.adsabs.harvard.edu/abs/2004PhRvD..70d1502G} {70, 041502}

\bibitem[\protect\citeauthoryear{{Hu} \& {Sugiyama}}{{Hu} \&
  {Sugiyama}}{1996}]{1996ApJ...471..542H}
{Hu} W.,  {Sugiyama} N.,  1996, \mn@doi [\apj] {10.1086/177989}, \href
  {https://ui.adsabs.harvard.edu/abs/1996ApJ...471..542H} {471, 542}

\bibitem[\protect\citeauthoryear{{Inman} \& {Ali-Ha{\"\i}moud}}{{Inman} \&
  {Ali-Ha{\"\i}moud}}{2019}]{2019PhRvD.100h3528I}
{Inman} D.,  {Ali-Ha{\"\i}moud} Y.,  2019, \mn@doi [\prd]
  {10.1103/PhysRevD.100.083528}, \href
  {https://ui.adsabs.harvard.edu/abs/2019PhRvD.100h3528I} {100, 083528}

\bibitem[\protect\citeauthoryear{{Inman} \& {Kohri}}{{Inman} \&
  {Kohri}}{2022}]{2022arXiv220714735I}
{Inman} D.,  {Kohri} K.,  2022, arXiv e-prints, \href
  {https://ui.adsabs.harvard.edu/abs/2022arXiv220714735I} {p. arXiv:2207.14735}

\bibitem[\protect\citeauthoryear{{Inomata}, {Kawasaki}, {Mukaida}  \&
  {Yanagida}}{{Inomata} et~al.}{2018}]{2018PhRvD..97d3514I}
{Inomata} K.,  {Kawasaki} M.,  {Mukaida} K.,   {Yanagida} T.~T.,  2018, \mn@doi
  [\prd] {10.1103/PhysRevD.97.043514}, \href
  {https://ui.adsabs.harvard.edu/abs/2018PhRvD..97d3514I} {97, 043514}

\bibitem[\protect\citeauthoryear{{Josan} \& {Green}}{{Josan} \&
  {Green}}{2010}]{2010PhRvD..82h3527J}
{Josan} A.~S.,  {Green} A.~M.,  2010, \mn@doi [\prd]
  {10.1103/PhysRevD.82.083527}, \href
  {https://ui.adsabs.harvard.edu/abs/2010PhRvD..82h3527J} {82, 083527}

\bibitem[\protect\citeauthoryear{{Kohri}, {Nakama}  \& {Suyama}}{{Kohri}
  et~al.}{2014}]{2014PhRvD..90h3514K}
{Kohri} K.,  {Nakama} T.,   {Suyama} T.,  2014, \mn@doi [\prd]
  {10.1103/PhysRevD.90.083514}, \href
  {https://ui.adsabs.harvard.edu/abs/2014PhRvD..90h3514K} {90, 083514}

\bibitem[\protect\citeauthoryear{{Kolb} \& {Tkachev}}{{Kolb} \&
  {Tkachev}}{1994}]{1994PhRvD..50..769K}
{Kolb} E.~W.,  {Tkachev} I.~I.,  1994, \mn@doi [\prd]
  {10.1103/PhysRevD.50.769}, \href
  {https://ui.adsabs.harvard.edu/abs/1994PhRvD..50..769K} {50, 769}

\bibitem[\protect\citeauthoryear{{Koushiappas} \& {Loeb}}{{Koushiappas} \&
  {Loeb}}{2017}]{2017PhRvL.119d1102K}
{Koushiappas} S.~M.,  {Loeb} A.,  2017, \mn@doi [\prl]
  {10.1103/PhysRevLett.119.041102}, \href
  {https://ui.adsabs.harvard.edu/abs/2017PhRvL.119d1102K} {119, 041102}

\bibitem[\protect\citeauthoryear{{Lacey} \& {Cole}}{{Lacey} \&
  {Cole}}{1993}]{1993MNRAS.262..627L}
{Lacey} C.,  {Cole} S.,  1993, \mn@doi [\mnras] {10.1093/mnras/262.3.627},
  \href {https://ui.adsabs.harvard.edu/abs/1993MNRAS.262..627L} {262, 627}

\bibitem[\protect\citeauthoryear{{Lacki} \& {Beacom}}{{Lacki} \&
  {Beacom}}{2010}]{2010ApJ...720L..67L}
{Lacki} B.~C.,  {Beacom} J.~F.,  2010, \mn@doi [\apjl]
  {10.1088/2041-8205/720/1/L67}, \href
  {https://ui.adsabs.harvard.edu/abs/2010ApJ...720L..67L} {720, L67}

\bibitem[\protect\citeauthoryear{{Loeb}}{{Loeb}}{2022}]{2022ApJ...929L..24L}
{Loeb} A.,  2022, \mn@doi [\apjl] {10.3847/2041-8213/ac6591}, \href
  {https://ui.adsabs.harvard.edu/abs/2022ApJ...929L..24L} {929, L24}

\bibitem[\protect\citeauthoryear{{Moradinezhad Dizgah}, {Franciolini}  \&
  {Riotto}}{{Moradinezhad Dizgah} et~al.}{2019}]{2019JCAP...11..001M}
{Moradinezhad Dizgah} A.,  {Franciolini} G.,   {Riotto} A.,  2019, \mn@doi
  [\jcap] {10.1088/1475-7516/2019/11/001}, \href
  {https://ui.adsabs.harvard.edu/abs/2019JCAP...11..001M} {2019, 001}

\bibitem[\protect\citeauthoryear{{Oguri}, {Takhistov}  \& {Kohri}}{{Oguri}
  et~al.}{2022}]{2022arXiv220805957O}
{Oguri} M.,  {Takhistov} V.,   {Kohri} K.,  2022, arXiv e-prints, \href
  {https://ui.adsabs.harvard.edu/abs/2022arXiv220805957O} {p. arXiv:2208.05957}

\bibitem[\protect\citeauthoryear{{Planck Collaboration} et~al.,}{{Planck
  Collaboration} et~al.}{2020}]{2020A&A...641A...6P}
{Planck Collaboration} et~al., 2020, \mn@doi [\aap]
  {10.1051/0004-6361/201833910}, \href
  {https://ui.adsabs.harvard.edu/abs/2020A&A...641A...6P} {641, A6}

\bibitem[\protect\citeauthoryear{{Raidal}, {Spethmann}, {Vaskonen}  \&
  {Veerm{\"a}e}}{{Raidal} et~al.}{2019}]{2019JCAP...02..018R}
{Raidal} M.,  {Spethmann} C.,  {Vaskonen} V.,   {Veerm{\"a}e} H.,  2019,
  \mn@doi [\jcap] {10.1088/1475-7516/2019/02/018}, \href
  {https://ui.adsabs.harvard.edu/abs/2019JCAP...02..018R} {2019, 018}

\bibitem[\protect\citeauthoryear{{Ricotti} \& {Gould}}{{Ricotti} \&
  {Gould}}{2009}]{2009ApJ...707..979R}
{Ricotti} M.,  {Gould} A.,  2009, \mn@doi [\apj] {10.1088/0004-637X/707/2/979},
  \href {https://ui.adsabs.harvard.edu/abs/2009ApJ...707..979R} {707, 979}

\bibitem[\protect\citeauthoryear{{Sasaki}, {Suyama}, {Tanaka}  \&
  {Yokoyama}}{{Sasaki} et~al.}{2016}]{2016PhRvL.117f1101S}
{Sasaki} M.,  {Suyama} T.,  {Tanaka} T.,   {Yokoyama} S.,  2016, \mn@doi [\prl]
  {10.1103/PhysRevLett.117.061101}, \href
  {https://ui.adsabs.harvard.edu/abs/2016PhRvL.117f1101S} {117, 061101}

\bibitem[\protect\citeauthoryear{{Sheth}, {Mo}  \& {Tormen}}{{Sheth}
  et~al.}{2001}]{2001MNRAS.323....1S}
{Sheth} R.~K.,  {Mo} H.~J.,   {Tormen} G.,  2001, \mn@doi [\mnras]
  {10.1046/j.1365-8711.2001.04006.x}, \href
  {https://ui.adsabs.harvard.edu/abs/2001MNRAS.323....1S} {323, 1}

\bibitem[\protect\citeauthoryear{{Silk}}{{Silk}}{2018}]{2018PhRvL.121w1105S}
{Silk} J.,  2018, \mn@doi [\prl] {10.1103/PhysRevLett.121.231105}, \href
  {https://ui.adsabs.harvard.edu/abs/2018PhRvL.121w1105S} {121, 231105}

\bibitem[\protect\citeauthoryear{{St{\"u}cker}, {Ogiya}, {Angulo},
  {Aguirre-Santaella}  \& {S{\'a}nchez-Conde}}{{St{\"u}cker}
  et~al.}{2022}]{2022arXiv220700604S}
{St{\"u}cker} J.,  {Ogiya} G.,  {Angulo} R.~E.,  {Aguirre-Santaella} A.,
  {S{\'a}nchez-Conde} M.~A.,  2022, arXiv e-prints, \href
  {https://ui.adsabs.harvard.edu/abs/2022arXiv220700604S} {p. arXiv:2207.00604}

\bibitem[\protect\citeauthoryear{{Volonteri}, {Habouzit}  \&
  {Colpi}}{{Volonteri} et~al.}{2021}]{2021NatRP...3..732V}
{Volonteri} M.,  {Habouzit} M.,   {Colpi} M.,  2021, \mn@doi [Nature Reviews
  Physics] {10.1038/s42254-021-00364-9}, \href
  {https://ui.adsabs.harvard.edu/abs/2021NatRP...3..732V} {3, 732}

\makeatother
\end{thebibliography}



\appendix

\section{Density of collapsed regions}\label{sec:collapsed}

\begin{figure}
	\centering
	\includegraphics[width=\columnwidth]{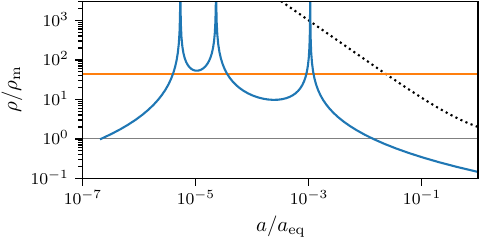}
	\caption{Density evolution for an isolated homogeneous ellipsoid with ellipticity $e=0.15$ during the radiation epoch. The ellipsoid's density $\rho$ (blue) significantly exceeds the background matter density $\rhoM$ (thin horizontal line) but does not approach the total density (dotted diagonal line), so it does not become gravitationally bound. Instead, after the collapse of each axis (spikes in the density), the ellipsoid's material drifts onward, bringing the density back down. In a more realistic picture, however, collapse of successively larger ellipsoids maintains a high density $\sim e^{-2}\rhoM$ (thick orange line) inside the collapsed region, leading to local matter domination, and hence halo formation, around $a\sim 0.02 \aeq$.}
	\label{fig:tophat}
\end{figure}

Equation~(\ref{density}) describes the density evolution of an isolated ellipsoidal top-hat overdensity both before and after collapse in the absence of gravitational forces. We illustrate this evolution in Fig.~\ref{fig:tophat}. For each of the three principal axes, the density increases until the axis collapses, after which it decreases again as particles drift back apart. In particular, the ellipsoid's density remains subdominant compared to the cosmic mean density (which includes radiation), so the no-gravity approximation remains valid. This picture suggests that most regions that collapse during the radiation epoch never form virialized structures. \citet{2013JCAP...11..059B} used a more mathematically precise argument to arrive at the same conclusion.

However, in a realistic scenario, each ellipsoid is embedded in a larger but less overdense ellipsoid. After the collapse of each axis, additional material in these larger ellipsoids begins to enter the first ellipsoid, raising its density. In particular, suppose successive ellipsoids $i$ have initial volumes proportional to $r_i^3$, so that the fraction of extra mass contributed by the $i$th ellipsoid beyond that already in the $(i-1)$th is $(r_{i}^3-r_{i-1}^3)/r_{i-1}^3$. Then one way to estimate the density of the collapsed region is to approximate that
\begin{equation}\label{rhosum}
    \frac{\rho}{\rhoM} = \frac{1}{3}\sum_i
    \frac{
    \theta(\lambda_1\delta_i\!-\!1)+\theta(\lambda_2\delta_i\!-\!1)+\theta(\lambda_3\delta_i\!-\!1)
    }{
    |1-\lambda_1\delta_i|\,|1-\lambda_2\delta_i|\,|1-\lambda_3\delta_i|
    }
    \frac{r_{i}^3\!-\!r_{i-1}^3}{r_{i-1}^3},
\end{equation}
where $\delta_i$ is the $i$th ellipsoid's enclosed linear density contrast and $\theta$ is the Heaviside step function. That is, we take each ellipsoid to contribute its density scaled by the fractional mass that it contributes to the system. The latter is just one third per collapsed axis of the total extra mass, $(r_{i}^3-r_{i-1}^3)/r_{i-1}^3$, a choice that is easy to justify if we imagine that the ellipsoids are instead initially cubical shells. Note that for simplicity, we also assume that all of the ellipsoids have the same axis ratios.

The linear density contrast $\delta$ is related to the primordial curvature fluctuation $\zeta$ by Eq.~(\ref{D}). In particular, we can write 
\begin{equation}\label{di}
    \delta_i = I_1 \zeta(r_i) \log(I_2 a/a_{H,i}),
\end{equation}
where the horizon scaling during radiation domination implies that $a_{H,i}=a_{H,0} r_i/r_0$. As an approximation, let us also write
\begin{equation}\label{zetapower}
    \zeta(r)=\zeta_0(r/r_0)^{-n}
\end{equation}
for some $\zeta_0$ and $n>0$. We expect that $n\simeq -\diff\log\Xi_\zeta/\diff\log r$, where $\Xi_\zeta$ is the volume-averaged correlation function
\begin{equation}
    \Xi_\zeta(r)=\frac{3}{r^3}\int_0^r r^{\prime 2}\diff r^\prime \xi_\zeta(r^\prime)=\int_0^\infty\frac{\diff k}{k}\mathcal{P_\zeta}(k) W(kr),
\end{equation}
where $\xi_\zeta(r)$ is the correlation function of $\zeta$, $\mathcal{P}_\zeta$ is the dimensionless power spectrum of $\zeta$, and $W(x)=(3/x^3)(\sin x-x\cos x)$ is the top-hat window function.

We can now take the continuum limit of Eq.~(\ref{rhosum}) to obtain an integral,
\begin{equation}
    \frac{\rho}{\rhoM} = \int\frac{\diff r}{r}
    \frac{
    \theta(\lambda_1\delta-1)+\theta(\lambda_2\delta-1)+\theta(\lambda_3\delta-1)
    }{
    (|1-\lambda_1\delta|+\epsilon)\,(|1-\lambda_2\delta|+\epsilon)\,(|1-\lambda_3\delta|+\epsilon)
    },
\end{equation}
where we add a small parameter $\epsilon>0$ to ensure convergence. Equations (\ref{di}) and~(\ref{zetapower}) imply
\begin{equation}
    \frac{\diff\log\delta}{\diff\log r}
    =
    -n - \log\left(I_2 \frac{a}{a_{H,0}}\frac{r_0}{r}\right)^{-1} \simeq -n,
\end{equation}
where the last approximation is valid in the late-time limit, long after the relevant modes have entered the horizon. In this limit,
\begin{align}\label{rhoint}
    \frac{\rho}{\rhoM} &\simeq \frac{1}{n}\int_0^\infty\frac{\diff \delta}{\delta}
    \frac{
    \theta(\lambda_1\delta-1)+\theta(\lambda_2\delta-1)+\theta(\lambda_3\delta-1)
    }{
    (|1\!-\!\lambda_1\delta|\!+\!\epsilon)\,(|1\!-\!\lambda_2\delta|\!+\!\epsilon)\,(|1\!-\!\lambda_3\delta|\!+\!\epsilon)
    }.
\end{align}
Recall that we define $\lambda_1=(1+3e+p)/3$, $\lambda_2=(1-2p)/3$, and $\lambda_3=(1-3e+p)/3$, so $\rho/\rhoM$ is a function of the ellipticity $e$, prolateness $p$, and $n$. This integral can be evaluated analytically, but the resulting expression is complicated. Instead, we comment that the outcome depends only weakly on $p$, which is typically close to zero, and is well approximated by
\begin{equation}\label{rhocoll}
    \frac{\rho}{\rhoM} \simeq \frac{\log_{10}(1/\epsilon)}{n e^2}.
\end{equation}

Roughly speaking, $\epsilon$ sets the minimum comoving separation of opposite ends of a collapsing ellipsoid, in units of their initial separation. Of course there is physically no minimum, but the resulting regularization of the density can be viewed as a consequence of the finite thickness of the initial phase sheet due to dark matter's thermal motion. If the dark matter is PBHs, it can be regarded instead as a consequence of the nonzero separation between individual particles. Alternatively, it may be viewed as a choice to neglect the formation of structures too far below the scales we are studying. In any event, for plausible choices of $\epsilon$, the prefactor in Eq.~(\ref{rhocoll}) is of order 1. The correlation slope $n$ is also typically of order 1, so $\rho/\rhoM\sim e^{-2}$ up to a factor of a few. We mark this density in Fig.~\ref{fig:tophat}. We also remark that if $\mathcal{P}_\zeta(k)$ is only a shallow function of $k$, then $n$ could be much smaller than 1, greatly boosting $\rho/\rhoM$.


\label{lastpage}
\end{document}